\documentclass[12pt,twoside,fleqn]{report}

\usepackage{latexsym,amssymb}

\usepackage[only,twltt,tentt,sevtt]{rawfonts}

\def\grg{$GRG_{\! EC}$}

\begin{document}

\begin{center}{\large
Specialized computer algebra system\\ 
for application in general relativity}\\[2ex]
S.I. Tertychniy
\end{center}
\nopagebreak

\begin{quote}\small 
{\bf Abstract:}
A 
brief characteristic 
of the 
specialized computer algebra system  \grg{}  
intended for symbolic computations in the field of
general relativity is given. 
\end{quote}

The code \grg{} 
constitute a full-fledged
programming system 
intended for application in the field of
the general
relativity and
adjacent areas of the differential
geometry and the classical field theory.
Written mostly in the lisp dialect known as {\sc standard lisp},
it 
is realized, structurally, as the top layer 
upon  the 
universal 
computer algebra system {\sl Reduce}.
The latter is 
utilized as the primary tool for
execution of the general kind symbolic
mathematical calculations.
The code infrastructure includes,
in particular, the user interface
based on the interpreter of
the 
so called {\it language of problem specification\/}
which
models 
the natural language
in its simplified version adapted to
the description of the notions and relationships 
taking place in
the application field.
The collection of 
algorithms implementing 
the set of {\it data objects\/}
and the rules of operations with them
models 
the most important notions and
relationships
(equations) 
established in the relevant 
areas of the physics and the geometry.
One could note in this respect
implementation of
the calculus of exterior forms,
the spinor algebra tools, 
the major elements of the tensor calculus.
(All these techniques operate with
separate object components, no abstract index methods
have been implemented).
The application specific algorithms enable one, 
in particular, to handle
various
bases in foliations of exterior forms 
connected with the metric structure,
the connection, 
the curvature with its irreducible constituents and
invariants,
the equations connecting the above objects
such as Cartan
equations, Bianchi equations, 
various algebraic identities, 
the field equations 
of the gravity theory (Einstein
equations).
The handling  of a number of the classical field
has been implemented including
electromagnetic field, 
massless spinor field, 
massive spinor fields,
massless scalar field, 
conformally invariant scalar
field, 
massive scalar field
and others.
It is worth noting also
the feasibility to manipulate with
Newman-Penrose spin coefficients, 
Lanczos representation of the conformal curvature, 
Rainich theory
of the coupling of electromagnetic and gravitational fields,
Killing vectors and more.

\noindent
\grg{} is currently 
 available free of charge at 
{\tt http://grg-ec.110mb.com}

\end{document}